\documentclass[preprint,aps] {revtex4}

\usepackage{graphicx}
\usepackage{dcolumn}
\usepackage{amsmath}
\usepackage{amssymb}
\usepackage{bm}
\usepackage{lscape}
\begin{document}
\title {
A method for estimating the cooperativity length in polymers
}
\author{Marco Pieruccini}
\affiliation{CNR, Istituto Nanoscienze, v. Campi 213/A, 41125 Modena, Italy}




\begin{abstract}
The problem of estimating the size of the cooperatively rearranging regions (CRRs) in supercooled polymeric melts from the analysis of the $\alpha$-process in ordinary relaxation experiments is addressed. The system is treated with the canonical formalism as an ensemble of CRRs, which are described by a stationary distribution relative to the rearrangement energy threshold. The process whereby a CRR changes its configuration is viewed as consisting of two distinct steps: a reduced number of monomers reaches initially an activated state allowing for some local rearrangement; then, the regression of the energy fluctuation may take place through the configurational degrees of freedom, thus allowing for further rearrangements on larger length-scales. The latter are indeed those to which the well known Donth's scheme refers. Two main regimes are envisaged, depending on wether the role played by the configurational degrees of freedom in the regression of the energy fluctuation is significant or not. It is argued that the latter case is related to the occurrence of an arrhenian dependence of the central relaxation time. Data of the literature are rediscussed within this new framework.
\end{abstract}

\maketitle

\subsection{Introduction}

Segmental relaxation in polymers, in particular on approaching the glass transition, is characterised by cooperativity. This is a condition whereby a monomer moves provided the sorrounding ones \emph{happen} to move concurrently. The role of chance here is fundamental and represents an important distinction from other kinds of motion, e.g. those involved in sound propagation.

The existence of an associated length scale of dynamical correlations is inherently addressed by this picture and was put farward already by Adam and Gibbs (A-G) in a seminal paper several decades ago, through the introduction of the concept of Cooperatively Rearranging Region (CRR)~\cite{A_G}. On approaching the dynamic glass transition temperature $T_g$ from above, this characteristic length would increase, and the problem of estimating its limiting value was considered by Donth in a number of papers (see e.g.~\cite{Donth_1}) and extensively treated in his book~\cite{Donth_book}. The relevance of this kind of motion, known as $\alpha$-process, thus has to do with the understanding of the glass transition.

Evidences of a growing length scale on decreasing the temperature have been reported as results of appropriate analyses of the $\alpha$-relaxation patterns observed by, e.g., ordinary dielectric or mechanical spectroscopies. Dielectric response analyzed in terms of either multipoint dynamic correlation functions~\cite{dalle} or Donth's approach~\cite{Saiter_polym}, and mechanical response analyzed within the framework of a canonical representation of an ensemble of cooperativity regions~\cite{TCA} (related to but not in general coincident with the CRRs, as explained below), all provide mutually consistent results, albeit from just a qualitative viewpoint. In particular, the average number $z$ of cooperatively rearranging units worked out from data within the canonical scheme appears to be much smaller than expected for a CRR.

On the contrary, when the same canonical approach is used to analyze data from samples crystallized at (almost) the same temperature where relaxation measurements were done, a natural \emph{thermodynamic} criterion for the arrest of crystallization emerges immediately~\cite{CPS}.

Further support to this model comes from the analysis of the relaxation in samples of poly(dimethyl siloxane) (PDMS) confined in nanoporous glasses~\cite{nanopores}.
Reducing the diameter of the pores, progressively hinders the glass transition (as indicated by a related lowering of $T_g$, accompanied also by a decrease of the specific heat step $\Delta c_p$) until it disappears at diameters $d_{pore}=5$~nm and below. A relaxation process following a Vogel Fulcher Tammann (VFT) dependence typical of a glassy dynamics, is revealed by dielectric spectroscopy for bulk and confined systems down to $d_{pore}=7.5$~nm. This dependence converts to arrhenian (proper of local modes) for $d_{pore}=5$~nm, and for $d_{pore}=2.5$~nm the relaxation totally disappears. On lowering the temperature and/or $d_{pore}$ the relaxation broadens; correspondingly, $z$ increases, as revealed by the analysis carried out with the canonical ensemble model~\cite{TCA} on the available data~\cite{nanopores}. Rearranging regions of the order of 1~nm size ($z\simeq 8$ at $T=136.6$~K) are thus found already at $d_{pore}=5$~nm, so that the absence of relaxation at $d_{pore}=2.5$~nm wouldn't be unexpected. It is worth noting that \emph{in the VFT regime} both dielectric and thermal spectroscopies provide quantitatively consistent results for the relaxation rates, that is, the dipole orientation fluctuations and the entropy fluctuations are coupled.

With regards to oriented systems, the canonical ensemble analysis of the segmental relaxation in cold-drawn semicrystalline poly(ethylene terephthalate) (PET), yields values for $z$~\cite{EPJE_2009} which may differ significantly from those estimated by a Donth-based analysis~\cite{Saiter_polym}.

The above arguments suggest that the cooperative dynamics considered in Donth's theory differs from that revealed by fitting data using the canonical ensemble approach (indeed, in 5~nm confined PDMS the latter "survives" while the glassy character of the motion disappears). Notwithstanding the differences, consistencies with experiments are found with both approaches; so, it might be argued that in fact the two schemes reveal different aspects of the same $\alpha$-process.

To be explicit at the outset, the following picture will be proposed. In the temperature interval where activated configurational motion sets in, i.e. between $T_g$ and a higher temperature $T_A$ around the melting point $T_m$ (if it exists), molecular units move about given locations~\cite{Lubchenko,biroli}, and only after a sufficient fluctuation they change cooperatively their configuration. This activated (local) rearrangement is a precursor for a subsequent, larger scale, cooperative motion wherever the energy fluctuation is allowed to regress through \emph{diffusive} conformational degrees of freedom (the attribute "diffusive" is adopted here as an easy way to refer to the configurational motion on larger length-scales, in contrast to the pre-transitional, oscillatory motion). The CRRs to which the Donth and A-G theories refer, are inherent to this large scale cooperativity. The fluctuation regression is not an activated process, and may take place also through other paths wherever the diffusive configurational motion is hindered (this might well be the case where relaxation takes place under particular confinement conditions, such as in the interlamellar regions of semicrystalline systems, or when chain orientation is overwhelmingly important).

The scope of the present paper is to provide a semi-phenomenological model for estimating the average size of the CRRs, which may be alternative (but not in contrast) to the approach of Donth. Furthermore, the scheme to be proposed below may suggest how the character of the $\alpha$-relaxation can happen to turn "local" when configurational constraints dominate (such as in the 5~nm confined PDMS). This model is not meant to compete with more fundamental approaches such as that of ref.~\cite{dalle}.

\section{A reminder on the canonical ensemble approach}

Consider a region consisting of $z$ molecular units in the polymeric melt. This subsystem is to be identified with the \emph{small} cooperativity region responsible of the local relaxation which may be observed also after the glassy dynamics has been suppressed. At variance from the original formulation~\cite{EPJE_2009}, the nature of this region is allowed to differ from that of the CRR introduced by A-G; in the present case, indeed, some limited mobility is allowed at its boundaries. Also here, however, all rearranging regions will still be taken to be the same size for simplicity.

Let $\zeta$ be the minimum energy per monomer to be gained by fluctuation for a rearrangement. As described in~\cite{EPJE_2009}, this threshold depends on the actual configuration of the region, so it fluctuates in time. After this barrier is crossed and a rearrangement initiated, all the energy initially gained is returned to the heat bath (this aspect was not given appropriate relevance in the original formulation). Before the transition takes place, the $z$ monomers undergo a collective motion exploring their actual basin of attraction in the phase space. A change of conformation requires that part of the energy $\left\langle E \right\rangle_\zeta$, i.e. roughly the average from the states \emph{above} the threshold $\zeta$, is gained by each of these monomers; its product with the probability $w(\zeta)$ that the monomer is in a rearranging state, yields approximately the actual amount that is absorbed to reach the top of the barrier. Once this activated state is attained, the configuration may change; however, since the energy threshold of the new configuration is known with just a probability $p(\zeta)$, the entropy $S = -k_B \int d\zeta\, p(\zeta)\,\ln p(\zeta)$ must be accounted for in describing this process from the statistical mechanical point of view. The physical meaning of $p$ is that of a distribution of monomers which are in a mobility state \emph{after} a barrier heigt $\zeta$ has been crossed. The entropy $S$ is thus related to the configurational changes of the (small) cooperativity regions and is central to the theory developed in ref.~\cite{EPJE_2009}.

The probability distribution $p(\zeta)$ is derived upon extremizing an appropriate potential; this yields
\begin{equation}\label{1}
	p(\zeta) \sim e^{-[w(\zeta) \left\langle E \right\rangle_\zeta + \lambda\Delta\mu(\zeta)]/k_B T} \,,
\end{equation}
where $k_B T$ is the thermal energy and $\lambda$ is a Lagrange multiplier related to the condition that the average $\overline{\Delta\mu} \equiv \int d\zeta\, p(\zeta)\,\Delta\mu(\zeta)$ is a constant; $\Delta\mu(\zeta)= -k_B T \ln w(\zeta)$ is the rearrangement chemical potential given in terms of the probability
\begin{equation}\label{2}
	w(\zeta) \equiv \frac{Z_{\zeta,n}}{Z_{0,n}} \,,
\end{equation}
which on its turn is expressed through the partition function
\begin{equation}\label{3}
	Z_{\zeta,n} \equiv \int^{\infty}_{\zeta} d\epsilon \, \epsilon ^n \, e^{-\epsilon/k_B T} \,.
\end{equation}
Note that, apart of a factor which is irrelevant presently, $Z_{0,n}$ is the partition function of an ensemble of $n+1$ independent oscillators~\cite{greiner}.

The meaning of eq.~\ref{1} is simple: The subset of (mobile) monomers with given $\zeta$ is larger the lower the collective rearrangement free energy $\lambda \Delta\mu(\zeta)$ is; on the other hand, the absorption of mean energy $w(\zeta) \left\langle E \right\rangle_\zeta$ promotes the transition to different configurations, i.e. to states with different $\zeta$ value, thus depleting the same subset.

\begin{figure}[h]
\includegraphics[width=8 cm, angle=0]{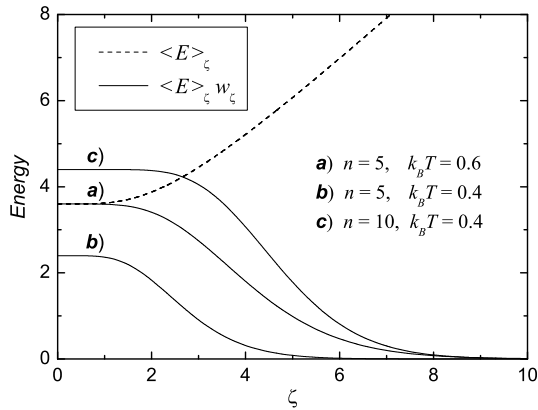}
\caption{\small{
Average energy $\left\langle E \right\rangle_\zeta$ and its product with $w_\zeta$ as functions of the energy threshold $\zeta$. All energies are expressed in kcal/mol.
}}
\label{fig1_art}
\end{figure}

Figure~\ref{fig1_art} shows the dependence on $\zeta$ of \emph{i}) the average energy $\left\langle E \right\rangle_\zeta$ and \emph{ii}) its product with the probability $w(\zeta)$, which can be expressed in terms of the incomplete gamma function as
\begin{equation}\label{4}
	w \left\langle E \right\rangle_\zeta = k_B T \,(n+1) \frac{\Gamma (n+2,\zeta/k_BT)}{\Gamma (n+2)} \,.
\end{equation}
Once $k_BT$ and $\zeta$ are assigned, a region can rearrange at a significant rate provided $n$ is large so as to make $w\left\langle E \right\rangle_\zeta > 0$ sufficiently. On increasing $\zeta$, also $n$ has to be incremented to maintain mobility. If the temperature lowers, on the other hand, the upper bound below which $w \left\langle E \right\rangle_\zeta$ is non-negligible shifts towards lower $\zeta$s and mobility is recovered once again assigning larger values to $n$. On fitting the data, $n$ is always found to increase with the number $z$ of units in a rearranging region, so the meaning of this argument is that the average size of the rearranging regions increases when either $T$ lowers at given $\zeta$ or the latter increases (by e.g. crosslinks or crystal confinement) at fixed $T$. In other words, mobility characterizes those regions for which $\overline{\Delta\mu}$ is appropriately small, and this requires that their size be large enough or $\zeta$ is low.

The relaxation function is an average over the distribution $p(\zeta)$ of single-time decays~\cite{EPJE_2009}:
\begin{equation}\label{5}
	\phi(t) \sim \int^{\infty}_{0} d\zeta \, w^\lambda \, e^{\,-w\left\langle E \right\rangle_\zeta} \, e^{-t/\tau(\zeta)} \,,
\end{equation}
where
\begin{equation}\label{6}
	\tau(\zeta) \equiv \tau^* e^{\,z\,\Delta\mu(\zeta)/k_B T}
\end{equation}
is the \emph{actual} relaxation time of the $z$ monomers in a configuration characterized by a barrier $\zeta$, and the identity $w^\lambda = \exp\{-\lambda\,\Delta\mu/k_B T\}$ has been used.
On fitting the data, one expects (as is found indeed) that $\lambda$ remains close to $z$ (i.e. within the same order); thus, the Lagrange multiplier associated to the condition $\overline{\Delta\mu} = const.$ during the transitions between configurations with different $\zeta$, is related to the number of units in the rearranging region.

\section{CRR size}

As anticipated in the introductory section, the onset of a \emph{large scale} cooperative rearrangement is viewed here as one among the possible paths through which an energy fluctuation of those considered above regresses. There are at least two aspects that should be taken into account regarding the relevance of this mechanism. On the one hand, a small enough specific heat step $\Delta c_p$ at $T_g$ should guarantee that the energy released by the activated $z$ monomers rapidly spreads out over a larger domain on the way back to the heat bath. On the other, the coupling to the diffusive configurational degrees of freedom should be efficient, which is expected indeed, since both the small and large scale motions considered relate to the same degrees of freedom.

\subsection{Rearrangement free energy vs. Configurational entropy}

As a preliminary step it is worth considering some known experimental results and their analysis with the canonical ensemble model. For convenience, the system to be considered specifically is PET.

Scanning temperature dynamic mechanical spectroscopy on an initially glassy PET sample shows a relatively narrow $\alpha$-process with a $\tan \delta$ peak centred at $T \approx 85\,^{\rm{o}}$C (probing frequency of 0.3~Hz and scanning rate of $2\,^{\rm{o}}$C/min)~\cite{Polymer_2003}. After cold-crystallization at $T_c = 100\,^{\rm{o}}$C, followed by an annealing at that temperature for an overall period of 7 hs, a similar analysis carried out in the same experimental conditions reveals a broader $\alpha$ peak centred at $T \approx 100\,^{\rm{o}}$C~\cite{EPJE_2008}. This is an effect of the dynamical constraints introduced by crystal formation.

Isothermal mechanical spectroscopy carried out on a semicrystalline PET sample, prepared as before from the same raw material, revealed a segmental relaxation process consisting of two components: a slow one, which is believed to characterize the interlamellar amorphous domains, and a fast one localized in amorphous pockets~\cite{Tiberio_PET} where dynamic constraints seem to be less effective. The analysis was carried out in the $10^{-3}-60$~Hz frequency range for temperatures of 85, 90 and $95\,^{\rm{o}}$C (for each temperature the frequency scan was repeated twice, showing no significant change in the loss pattern). Details can be found in ref.~\cite{Polymer_2011}, but the relevant outcomes for the present discussion are the fast mode $\overline{\Delta\mu}$ values worked out by means of the canonical ensemble scheme and reported in Table~\ref{tab1} for easy reference (the data of ref.~\cite{Polymer_2011} have been re-analyzed here with better integration routines).
\begin{table}[h]
\begin{tabular}{ccc}
\hline
  		 				 	&		                     &						\\
       $T$      & $\overline{\Delta\mu}$ &  $T\,s_c$  \\
  		 				 	&		                     &						\\
\hline
							  &			                   &		        \\
	     85	      &    ~~~~~ 0.67 ~~~~~    &    0.77    \\
							  &			                   &		        \\
	     90	      &          1.00          &    0.87    \\
							  &			                   &		        \\
	     95	      &          1.36          &    1.00    \\
							  &			                   &		        \\
\hline
\end{tabular}

\caption{Rearrangement chemical potential $\overline{\Delta\mu}$ and configurational entropy per monomer $s_c$ times $T$ for semicrystalline PET (both in kcal/mol) at different temperatures (in $^{\rm{o}}$C), calculated for a Kauzmann temperature $T_K=318.7$~K and a specific heat step per monomer $\Delta c_p\approx \, 9.3\, k_B$ at $T_g$.}
\label{tab1}
\end{table}

The central relaxation times of the fast component have been fitted with the VFT function
\begin{equation}\label{VFT}
	\tau_0 = \tau_\infty e^{D T_{\scriptscriptstyle{\rm{VFT}}}/(T-T_{\scriptscriptstyle{\rm{VFT}}})} \,,
\end{equation}
with $\tau_\infty$ set equal to $10^{-14}$~s, yielding $T_{\scriptscriptstyle{\rm{VFT}}} = 318.7$~K~(cf. \cite{Polymer_2011} within errors). This value is in agreement with independent analyses~\cite{Saiter_polym} and has been identified with the Kauzmann temperature $T_K$ for an estimate of the specific configurational entropy
\begin{equation}\label{7}
	s_c = \Delta c_p \ln \left(\frac{T}{T_K}\right)
\end{equation}
The product $Ts_c$ is reported in Table~\ref{tab1} as well, and it can be readily noticed that the rearrangement chemical potential $\overline{\Delta\mu}$ is larger than $Ts_c$ except for $T$ close to the lowest value. Given the above results from dynamic mechanical spectroscopy, and being $T_g \approx 75\,^{\rm{o}}$C for amorphous PET, a value of $85\,^{\rm{o}}$C can be considered to be very close to the glass transition temperature within the amorphous pockets of the crystallized sample.

The data listed in the table suggest that the approximate relation
\begin{equation}\label{8}
	\frac{\overline{\Delta\mu}}{T} \gtrsim s_c
\end{equation}
(to be implemented below in a more appropriate form) may relate to the actual criterion for the onset of large scale mobility. The configurational entropy would play the role of a rearrangement threshold. Below $T_g$, small scale fluctuations may occur, but are unable to excite significant diffusive motions upon regression, so the energy is returned to the heat bath through different paths and a structural arrest is observed.

The meaning of eq.~\ref{8} becomes clear if it is re-expressed in terms of probabilities. Let $\tilde{\zeta}$ be a characteristic value of the energy threshold for which $\overline{\Delta\mu} = -k_B T \ln (Z_{\tilde{\zeta},n}/Z_{0,n})$, and let $\Omega_c$ be the number of (low energy) configurational states counted on a \emph{per monomer} basis, such that $\Omega_c \sim e^{\,s_c/k_B}$. Then, eq.~\ref{8} reads
\begin{equation}\label{8_bis}
	\Omega_c^{-1}\,\gtrsim\,\frac{Z_{\tilde{\zeta},n}}{Z_{0,n}} \,,
\end{equation}
that is, the probability associated to the final state is larger than that of the initial activated state; in other words, the fluctuation regression is spontaneous.

Wherever the energy needed to induce a configurational change is too large (e.g. in the 5~nm confined PDMS of the introduction), there is no possibility to get it from the activated state; the fluctuation energy returns to the heat bath through other mechanisms and the number of rearranging units remains $z$. The process, then, is (almost) local.

\subsection{Large scale cooperativity}

Based on the suggestion given by eq.~\ref{8}, it is assumed as a working hypothesis that all the energy of the activated regions for which $\Delta\mu (\zeta) > T s_c$, is completely delivered to the diffusive conformational degrees of freedom when the fluctuation regresses (this models the energy transfer efficiency considered at the beginning of the section). This energy induces, on a number $N_{ind}$ of surrounding monomers, \emph{actual} configurational motion with an associated entropy of $N_{ind}\,s_c$. Consistently with an efficiency assumed to be high, the energy transfer to the $N_{ind}$ monomers is considered to be approximately reversible, yielding
\begin{equation}\label{9}
	N_{ind} \approx z \frac{\kappa \overline{\zeta} }{Ts_c} \,,
\end{equation}
where
\begin{equation}\label{10}
	\kappa\overline{\zeta} = \frac{1}{M} \int^{\infty}_{\zeta_{0}} d\zeta\,\zeta\,p(\zeta)
\end{equation}
is the fraction $\kappa = M^{-1} \int^{\infty}_{\zeta_{0}} d\zeta\,p(\zeta)$ of mobile regions with $\Delta\mu (\zeta)> s_c$, times the average energy associated to each of its $z$ monomers; $M\equiv \int^{\infty}_{0}d\zeta\,p$ is the normalization constant for $p$ and $\zeta_{0}$ is such that $\Delta\mu (\zeta_0) = T s_c$. Then, the total number of monomers which actually rearrange cooperatively is given by
\begin{equation}\label{11}
	N_\alpha = N_{ind} + z \,.
\end{equation}

As it can be imagined, eq.~\ref{8} does not have to be taken in a strict sense as a condition for large scale rearrangement, since indeed it is an indicative criterion involving averages. The relevant quantities are in fact the mean energy $\overline{\zeta}$ of the fraction $\kappa$ of monomers above the threshold, and these may be significant also in situations where eq.~\ref{8} is not, or marginally not fulfilled.

\section{Analysis of experimental results}

In order to illustrate some practical application of eq.~\ref{9}, PET will be considered again in some detail. This choice follows from the fact that this material is a good model system for the investigation of basic polymer physics; moreover, a number of experimental data extracted under a variety of conditions is available to the author.

In general, one starts from the Havriliak-Negami (H-N) analysis of mechanical or dielectric losses determined in isothermal conditions for an as wide as possible frequency range. Details can be found, e.g., in refs.~\cite{TCA,Polymer_2011,EPJE_2009}. Once the H-N parameters for the process of interest have been found, the corresponding relaxation function $\phi_{HN}(t)$ is calculated by cosine Fourier transforming:
\begin{equation}\label{12}
	\phi_{HN}(t) = \frac{2}{\pi}\int^{\infty}_{0}\frac{ A ''(\omega; a, b, \tau_0)}{\Delta A}\cos (\omega t)\frac{d\omega}{\omega} \,,
\end{equation}
where $A''$ is the imaginary part of the complex response function (i.e. either the dielectric permittivity or the mechanical modulus) expressed in the form
\begin{equation}\label{13}
	A = \frac{\Delta A}{\left[1 + (i \omega\tau_0)^a\right]^b} \,,
\end{equation}
with $a$ and $b$ the width and asymmetry parameters ($a,b\leq1$; $b=1$ for a symmetric process) and $\tau_0$ the central relaxation time.

The analysis then proceeds by adjusting eq.~\ref{5} on the experimental relaxation function $\phi_{HN}(t)$. The fitting parameters are $\lambda$, $z$ and $\tau^*$, while the exponent $n$ is chosen in order to reach a minimum $\chi^2$ with the prescription that the lower limit $t_{min}$ of the fitting interval is pushed towards low $t$-values so to keep it close to $\tau^*$.

With this procedure, the average rearrangement chemical potential $\overline{\Delta\mu}$, the average height of the energy threshold $\left\langle \zeta\right\rangle$ and all other ingredients needed to calculate $N_\alpha$ can be derived. $\Delta c_p$ is obtained by calorimetry and $T_K$ is identified with $T_{\scriptscriptstyle{\rm{VFT}}}$ in fitting $\tau_0 (T)$ with eq.~\ref{VFT}.

After each fitting, it is always found that $\lambda$ and $z$ are of the same order and
\begin{equation}\label{15}
	\tau_0 \approx \tau^*\, e^{\,\lambda\overline{\Delta\mu}/k_B T} \,.
\end{equation}
The fulfillment of these conditions reflects a non-independence of the fitting parameters, which represents an effective reduction of their number.

\subsection{$100\,^{\rm{o}}$C crystallized PET}

\subsubsection{Before re-crystallization}

\begin{figure}[h]
\includegraphics[width=8 cm, angle=0]{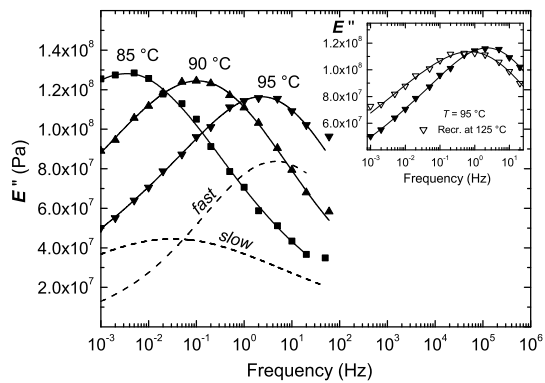}
\caption{\small{
Mechanical loss patterns of a PET sample cold-crystallized at $T_c =100\,^{\rm{o}}$C for 7 h. The dashed lines represent the fast and slow components deconvoluted from the $T =95\,^{\rm{o}}$C curve. The inset shows a comparison of the responses before and after re-crystallization at $125\,^{\rm{o}}$C for 2 h.
}}
\label{fig2_art}
\end{figure}

The dynamic mechanical analysis of a sample cold-crystallized at $T_c =100\,^{\rm{o}}$C and annealed at the same temperature for an overall duration of 7 h, yields the loss patterns of Fig.~\ref{fig2_art} relative to the temperatures of 85, 90 and 95$\,^{\rm{o}}$C.

For $T=90$ and $95\,^{\rm{o}}$C it is possible to separate the fast and slow contributions as shown in the figure (at $85\,^{\rm{o}}$C the slow component appears to be very weak in the observable frequency interval). For a comparison with previous analyses carried out under the scheme of Donth, just the fast mode relaxation must be considered. Contributions to the entropy fluctuations arising from the slow mode are also present, but are difficult to estimate because of the lack of an available, wide enough, analysis in temperature (e.g. to get an estimate of the appropriate $T_K$). In any case this effect is expected to be small, due to the highly constrained dynamics characterizing this relaxation process.

Table~\ref{tab2} collects the results of the fitting procedure to the responses of Fig.~\ref{fig2_art} (note that $\overline{\Delta\mu} \sim \delta[\Delta\mu]$ is found, which is characteristic of a thermodynamically small system~\cite{EPJE_2009}). Assuming $T\approx 85\,^{\rm{o}}$C to be the temperature at which the relaxation dynamics within the amorphous pockets resemble at best the relaxation dynamics of completely amorphous PET at $T_g$, it turns out that the present estimates of $N_\alpha$ agree fairly well with those provided in ref.~\cite{Saiter_polym}, i.e. $\sim 100$, $75$ and $50$ at the corresponding temperatures considered here.

Estimating the CRR diameter with the expression
\begin{equation}
	\xi \sim \sqrt[3]{\frac{W N_{\alpha}}{\rho N_A}} \,,
\end{equation}
with $W$ the molecular weight of the monomer, $\rho$ the density and $N_A$ the Avogadro number, one finds it around 2.8~nm, 2.2~nm and 1.8~nm respectively at 85, 90 and 95$\,^{\rm{o}}$C.

\begin{table}[h]
    \begin{center}
        \begin{tabular}{ccccccccccccc}
        \hline
      &     &          &     &           &     &          &           &         &            &            &          & 						\\
  $T$ & $a$ & $\tau_0$ & ~~$n$~~ & $\lambda$ & ~~$z$~~~ & $\tau^*$ & $t_{min}$ & $\left\langle \zeta\right\rangle$ & $\overline{\Delta\mu}$ & $\delta[\Delta\mu]$ & $\kappa$ & $N_\alpha$  \\
 ($^{\rm{o}}$C) & &  (s)  &  &           &     &   (s)    &     (s)   &(kcal/mol)&  (kcal/mol) & (kcal/mol) &          &             \\
                & &       &  &           &     &          &           &          &             &            &          &	  				 \\
        \hline
  
  85 & 0.31  &    50 &   18  &  9.8  & 9.8  & $2.3\times 10^{-3}$ & $2.6\times 10^{-3}$ & 13.4 & 0.67 & 0.36 & 0.39 &  96		      \\
  90 & 0.34  &  0.64 &   10  &   4   & 5.8  & $1.2\times 10^{-4}$ & $2.4\times 10^{-4}$ &  9.1 &   1  & 0.56 & 0.63 &  49		      \\
  95 & 0.37  & 0.035 &    7  &  2.3  & 3.9  & $2.5\times 10^{-5}$ & $2.6\times 10^{-5}$ &  7.6 & 1.36 & 0.76 & 0.75 &  28		      \\
        
        \hline
        \end{tabular}
    \end{center}
    
    \caption{Temperature $T$, H-N parameters $a$ and $\tau_0$ ($b=1$ because the profile is symmetric, as a significant amount of crystals is present), fitting parameters $\lambda$, $z$, $\tau^*$ (and $n$), lower limit of the fitting interval $t_{min}$, and resulting values for average energy threshold $\left\langle \zeta\right\rangle$, rearrangement chemical potential $\overline{\Delta\mu}$ and its dispersion around the mean $\delta[\Delta\mu]$, fraction $\kappa$ of CRRs inducing diffusive configurational changes and total number of rearranging monomers $N_\alpha$, for the fast relaxation process of the $T_c =100\,^{\rm{o}}$C cold-crystallized PET sample at the temperatures of 85, 90 and 95$\,^{\rm{o}}$C.
     }
    \label{tab2}
\end{table}


\subsubsection{After recrystallization}

Semicrystalline samples prepared like those considered above, have been afterwards recrystallized for 2 h at the temperatures of either 115 or 125$\,^{\rm{o}}$C, i.e. just below and right above the annealing peak~\cite{Polymer_2008}. The pattern collected at 95$\,^{\rm{o}}$C after recrystallization at 125$\,^{\rm{o}}$C is compared with that obtained before recrystallization in the inset of Fig.~\ref{fig2_art}. The downwards shift of the frequency profile denotes a slowing down of the dynamics; a similar shift is observed after a 115$\,^{\rm{o}}$C recrystallization.

The detailed results of relaxation analyses carried out as before, are not reported (cf. ref.~\cite{Polymer_2011}); rather, some relevant worked out parameters are listed in Table~\ref{tab3}.

\begin{table}[h]
\begin{tabular}{lcccccc}
\hline
		 	       &             &		                       &		        &              &            &            \\
  $T$/sample & ~~$\left\langle \zeta\right\rangle$~~ &~~$\overline{\Delta\mu}$~~ &~~$\kappa$~~&~~$N_\alpha$~~& ~~$z$~~  &~~$\tau_0$~~\\ 
  				 	 &		         &                           &						&              &            &            \\
\hline
 				     &		          &                          &						&              &            &            \\
 90 n-r   	 &		  9.1     &            1             &		0.63		&      49      &    5.8     &    0.64    \\
 90 r-115    &		   9      &           1.02           &		0.60		&      44      &    5.4     &    2.01    \\
 90 r-125    &		 11.7     &           0.74           &		0.38		&      68      &    9.6     &    3.2     \\
 		         &		          &                          &						&              &            &            \\
\hline
 				     &		          &                          &						&              &            &            \\
 95 n-r   	 &		  7.6     &           1.36           &		0.75		&      28      &    3.9     &   0.035    \\
 95 r-115    &		  7.3     &           1.24           &		0.62		&      28      &    4.4     &   0.084    \\
 95 r-125    &		   7      &           1.16           &		0.58		&      30      &    4.9     &   0.07     \\
 		         &		          &                          &						&              &            &            \\
\hline
\end{tabular}

\caption{Average fluctuation energy $\left\langle \zeta\right\rangle$ and rearrangement chemical potential $\overline{\Delta\mu}$ (both in kcal/mol), fraction $\kappa$ of CRRs delivering energy to the diffusional configuration degrees of freedom, overall number $N_\alpha$ of readjusting monomers and central relaxation time $\tau_0$ (in s) of the fast mode for the samples indicated. In the leftmost column the number refers to the measurement temperature expressed in $^{\rm{o}}$C, while "n-r", "r-115" and "r-125" stand for non-recrystallized, recrystallized at 115$\,^{\rm{o}}$C and recrystallized at 125$\,^{\rm{o}}$C respectively.}
\label{tab3}
\end{table}

The fraction of material within the amorphous pockets can be estimated as $\eta = 1-\alpha/\alpha_L$, where $\alpha$ is the overall crystallinity of the sample, and $\alpha_L$ is the \emph{linear} crystallinity characterizing the lamellar stacks. From the data collected in ref.~\cite{mecmat} it is found that recrystallization at 115$\,^{\rm{o}}$C slightly increases $\eta$ from 0.39 to 0.42; but it is the recrystallization at 125$\,^{\rm{o}}$C which causes a most dramatic structural change, accompanied by a decrease of $\eta$ down to 0.26. It is impressive to observe how closely $N_\alpha$, as estimated at a temperature of $90\,^{\rm{o}}$C, follows this trend. (Note that $N_\alpha$ is underestimated for the recrystallized samples, because the value $T_K= 318.7$~K has been used for its calculation, while it should be slightly larger~\cite{Polymer_2008}.)  At the higher temperature of 95$\,^{\rm{o}}$C, instead, the size of the activated rearranging region (i.e. $z$) is small enough not to feel the effect of the constraints at the borders, and also the overall CRR size remains almost insensitive to recrystallization. These results point to a significant reduction in size of the amorphous pockets, as a result of restructuring above the annealing peak.

In this case, the enhancement of configurational constraints leads to an increase of cooperativity and of mean relaxation time.

\subsection{$160\,^{\rm{o}}$C crystallized PET}

\begin{figure}[h]
\includegraphics[width=8 cm, angle=0]{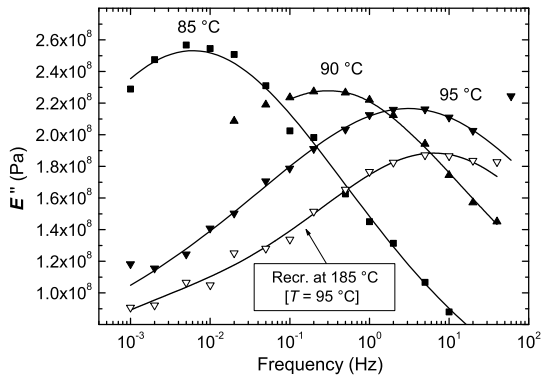}
\caption{\small{
Mechanical loss patterns of a PET sample cold-crystallized at $T_c =160\,^{\rm{o}}$C for 7 h. In the same frame is shown the response after recrystallization for 2 h at $185\,^{\rm{o}}$C, the measurement was done at a temperature of $95\,^{\rm{o}}$C.
}}
\label{fig3_art}
\end{figure}

Starting from the same raw material, semicrystalline PET samples have been prepared as before, but the temperature for crystallization and annealing was rised up at $T_c =160\,^{\rm{o}}$C. The mechanical loss patterns are shown in Fig.~\ref{fig3_art}; however, the value $T_{\scriptscriptstyle{\rm{VFT}}}\simeq 345$~K found seems too large (this is probably a consequence of the indeterminacy deriving from the rather broad maximum of the available $90\,^{\rm{o}}$C pattern; the deconvolution of the fast component couldn't thus be more then guessed), so the orientative value of $T_K=330$~K was adopted (consider that, for the $100\,^{\rm{o}}$C sample recrystallized above the annealing peak, a value of $\sim328$~K was found~\cite{Polymer_2011}). Table~\ref{tab4} reports the relevant quantities worked out with the relaxation fitting procedure used before.

\begin{table}[h]
\begin{tabular}{lccccccc}
\hline
		 	       &               &		                       &		        &              &            &           &          \\
  $T$/sample & ~~$\left\langle \zeta\right\rangle$~~ &~~$\overline{\Delta\mu}$~~ &~~$\kappa$~~&~~$N_\alpha$~~&   ~~$z$~~  &~~~~$\tau_0$~~~~& $\eta$ \\ 
  				 	 &		           &                           &						&              &            &           &          \\
\hline
 				     &		          &                          &						&              &            &            &          \\
 95 n-r   	 &		  7.5     &           1.05           &		0.68		&      55      &    6.1     &   0.035    &   0.25   \\
 95 r-175    &		  6.7     &           1.35           &		0.77		&      33      &    3.6     &   0.02     &   0.24   \\
 95 r-185    &		  6.8     &           1.4            &		0.78		&      33      &    3.6     &   0.015    &   0.14   \\
 		         &		          &                          &						&              &            &            &          \\
\hline
\end{tabular}

\caption{
Same as Table~\ref{tab3} for the sample crystallized at 160$\,^{\rm{o}}$C for 7 h and subsequently recrystallized at either 175 or 185$\,^{\rm{o}}$C. In the leftmost column, "n-r", "r-175" and "r-185" stand for non-recrystallized, recrystallized at 175$\,^{\rm{o}}$C and recrystallized at 185$\,^{\rm{o}}$C respectively. The parameter $\eta$ defined in the text is also reported.
}
\label{tab4}
\end{table}

With regards to the fast relaxation mode, thus, recrystallization at 185$\,^{\rm{o}}$C, i.e. above the annealing peak, does not yield significantly different changes with respect to the 175$\,^{\rm{o}}$C recrystallization, although an appreciable restructuring can be argued from the change in $\eta$, accompanied by a significant reduction of $N_\alpha$.

\subsection{Oriented PET}

The analysis of cold-drawn PET (drawing ratio of 4), subsequently crystallized at 140$\,^{\rm{o}}$C yields a number of rearranging units (at the small scale) of $z\simeq 7$ at $T=130\,^{\rm{o}}$C~\cite{EPJE_2009}. This result is in the order of the number of rearranging units recently estimated for semicrystalline PET samples with a similar drawing ratio (but different thermal history), on the basis of Donth's approach~\cite{Saiter_polym}. However, a thorough comparison of the two methods for the calculation of the cooperativity should be carried out. For a drawing ratio of 6, however, the number of rearranging units appears to be much lower.

It is worth noting that in these oriented PET samples the temperature dependence of the segmental relaxation is markedly arrhenian~\cite{Saiter_polym,JCP_2007}, pointing to a local character of the process. In this case eq.~\ref{9} does not apply; it is expected that the large scale configurational motion is hindered and, thus, irrelevant for the regression of the energy fluctuation.

\section{Concluding remarks}

The approach offered in the present scheme for the estimate of the (large scale) cooperativity length, differs significantly from that developed by Donth. In the latter, entropy or temperature fluctuations are directly accessed experimentally and the characteristic length of cooperativity calculated. On other hand, the same quantity is estimated here on the basis of different data sets; indeed the analysis of the relaxation focuses on the (local) activation process precursory to the configurational change and resolved in frequency for a fixed temperature. It is thus remarkable that the two methods provide mutually consistent results (see the case of low-$T_c$ crystallized PET).

It is found that, in the regime of glassy dynamics, the introduction of dynamical constraints causes an increase of $\left\langle \zeta\right\rangle$ and $z$, which is in general accompanied by a corresponding increase of $N_\alpha$. This is observed when low-$T_c$ PET recrystallization is performed and when PDMS is confined in pores of decreasing size. The increase in both $\left\langle \zeta\right\rangle$ and $z$ is connected with a corresponding enhancement of the relaxation width~\cite{TCA}.

The depletion of the constraints observed in the high-$T_c$ recrystallized PET samples, together with the corresponding acceleration of the dynamics, conforms to the picture of glass transition hindering proposed by Donth~\cite{Donth-Schick}. Yet, the real mechanisms determining this occurrence, and the reasons why so different behaviors are observed, still remain unknown. Whereas in recrystallized PET, acceleration or slowing down of the dynamcs is accompanied by a respective narrowing or broadening of the frequency profile, in the case of confined PDMS the acceleration is accompanied by a broadening in the relaxation profile. This broadening would be consistent~\cite{TCA} with the presence of a rigid layer forming at the borders of the PDMS-filled regions~\cite{nanopores}, so the acceleration mechanism is found to be related with a decrease of $\tau^*$. Modelling appropriately this fitting parameter, also including the effect of long range configurational fluctuations, would certainly help in the disentanglement of this problem.

Related to last point, also another aspect has to be considered. In fact, the analysis presented above starts from the assessment that the system is already in a regime of glassy dynamics, and only then the estimate of the large scale cooperativity is done. In the case of PDMS confined in 5~nm pores the $T$ dependence of the central relaxation time is found to be arrhenian, there appears to be no finite $T_K$ and the calculation of cooperativity is limited to the small scale region; afterwards, the size of these regions is found to be consistent with the geometrical constraints that have been imposed.

The problem of cooperativity is cast here in another light, allowing for familiar ideas to be recalled for its treatment; moreover, the present formulation highlights some aspects which may be worth considering when cooperativity has to be estimated under particular conditions, such as in strong confinement. Of course it is a model, and as such, it calls for a better structured theoretical framework and/or further support from the experiments.

\end{document}